\documentclass{nature}
\title{The Dipole Repeller}
\author{Yehuda Hoffman$^1$, Daniel Pomar\`ede$^2$, R. Brent Tully$^{3}$ and H\'el\`ene Courtois$^4$}

\usepackage{color}
\usepackage{graphics}
\usepackage[pdftex]{graphicx}
\usepackage{epstopdf}
\usepackage{epsfig}

\def  \LCDM{$\Lambda$CDM}

\def \kms {{\rm km s$^{-1}$}}

\def\VbulkR{{\bf V}$_{\rm bulk}$(R)}

\begin{document}
\maketitle

{\it In accordance with Nature publication policy, this version is the original one submitted to Nature Astronomy.
The accepted version is accessible online at Nature Astronomy and differs from this one. 
Nature Astronomy 2017, Volume 1 , article 36.
We will be allowed to post in 6 months the accepted version on arxiv.\\}

\begin{affiliations}
 \item Racah Institute of Physics, Hebrew University, Jerusalem 91904, Israel
 \item Institut de Recherche sur les Lois Fondamentales de l'Univers, CEA, Universit\'eŽ Paris-Saclay, 91191 Gif-sur-Yvette, France
 \item Institute for Astronomy (IFA), University of Hawaii, 2680 Woodlawn Drive, HI 96822, USA
 \item University of Lyon; UCB Lyon 1/CNRS/IN2P3; IPN Lyon, France
 \end{affiliations}

\begin{abstract}
In the standard (\LCDM)  model of cosmology the   universe  has emerged out of an early homogeneous and isotropic phase. Structure formation  is associated with the growth of density irregularities and   peculiar velocities.  
Our  Local Group  is moving with respect to the cosmic microwave background   (CMB) with a velocity of $V_{\rm CMB}= 631 \pm 20$~\kms\   \cite{1996ApJ...473..576F} and participates in a bulk flow that extends out to   distances of at least     
$\approx 20,000$~\kms\ \cite{2011ApJ...736...93N,2015MNRAS.447..132W,2015MNRAS.449.4494H}. The quest for the sources of that motion  has dominated cosmography since the discovery of the CMB dipole. 
The implicit assumption was   that excesses in the abundance of galaxies induce the Local Group motion\cite{1986ApJ...307...91L,1988ApJ...326...19L,1991Natur.350..391D}.
 Yet, underdense regions push as much as overdensities attract\cite{1988MNRAS.234..677L} but they are deficient of  light and consequently  difficult to chart. 
 It was suggested a decade  ago that an underdensity  in the northern hemisphere roughly 15,000~\kms\ away is a significan factor in the local flow\cite{2006ApJ...645.1043K}. 
 Here we report on kinematic evidence for such an underdensity. We  map  the large scale  3D velocity field  using a Wiener filter reconstruction from the  Cosmicflows-2 dataset of peculiar velocities, and identify  the attractors and repellers that dominate the local dynamics. 
We show here that the local   flow is dominated by a single attractor - associated with the Shapley Concentration - and a single previously unidentified repeller.  
Multipole expansion of the local flow provides further support for the existence and role played by the attractor and repeller.
The  bulk flow (i.e. dipole moment) is closely   (anti)aligned with the  repeller at  a distance of $16,000 \pm 4,500 $ ~\kms. The expansion eigenvector of the  shear tensor  (i.e. quadrupole moment) is closely aligned with the Shapley Attractor out to $\approx$7,000~\kms. The close alignment of the local bulk flow with the repeller provides further support for its dominant role   in shaping the local flow.
This Dipole Repeller is predicted to be associated with a void in the distribution of galaxies.  
\end{abstract}

The large scale structure of the universe is encoded in the flow field of galaxies.  A detailed analysis of the flow  uncovers the rich structure manifested by the distribution of galaxies, such as the prominent nearby clusters\cite{1990ApJ...364..349D,1999ApJ...520..413Z,2012ApJ...744...43C,2013AJ....146...69C}, the Laniakea  supercluster\cite{2014Natur.513...71T} and the Arrowhead mini-supercluster\cite{2015ApJ...812...17P}.  
A one-to-one correspondence  between the observed density field, derived from redshift surveys, and the reconstructed 3D flow field  has been established out to beyond 100 megaparsecs and down to a resolution of a  few megaparsecs\cite{2013AJ....146...69C}. Yet, the flow  contains more information on distant structures from tides and from continuity across the zone obscured by the Galactic disk, the so-called Zone of Avoidance\cite{1999ApJ...520..413Z,2001astro.ph..2190H}.  
 The Cosmicflows-2 dataset of peculiar velocities\cite{2013AJ....146...86T} provides reasonably dense coverage to $R \approx 10,000$~\kms\ (distances are expressed in terms of their equivalent Hubble velocity).
   
However convergent features in the large scale flow patterns reveal important influences at $R\sim 16,000$~\kms\ at the extremity of the data coverage.

The  linear 3D velocity field is reconstructed  here from the Cosmicflows-2 data by  the Bayesian methodology of the Wiener Filter (WF) and constrained realizations (CRs; see Methods). 
The WF  is a Bayesian estimator which assumes a prior model - here it is the $\Lambda$CDM model. 
It is a conservative estimator which balances between the data and its errors and the assumed prior model. Where the data is weak the WF estimation tends to the null hypothesis of a homogenous universe. The variance around the mean WF estimator is sampled by the CRs.

The WF   is used to construct the cosmic velocity and the cosmic web.  
The web is defined here by the velocity shear tensor\cite{2012MNRAS.425.2049H} - the web elements of the so-called V-web are defined by the number of eigenvalues of the tensor above a threshold value   (see   Methods). In the linear regime the flow  is irrotational and  constitutes a gradient of a  scalar potential.
Figure 1 shows the  large scale structure out to a distance of 16,000~\kms. Three different aspects of the flow  are depicted: streamlines which manifest the direction and magnitude of the velocity field (but do not represent trajectories; see  Methods), red and grey surfaces present the knots and filaments of the V-web 
 and the green-yellow surfaces correspond to the velocity potential.

\begin{figure}
\label{fig1}
\includegraphics[width=1.\textwidth,angle=-00]{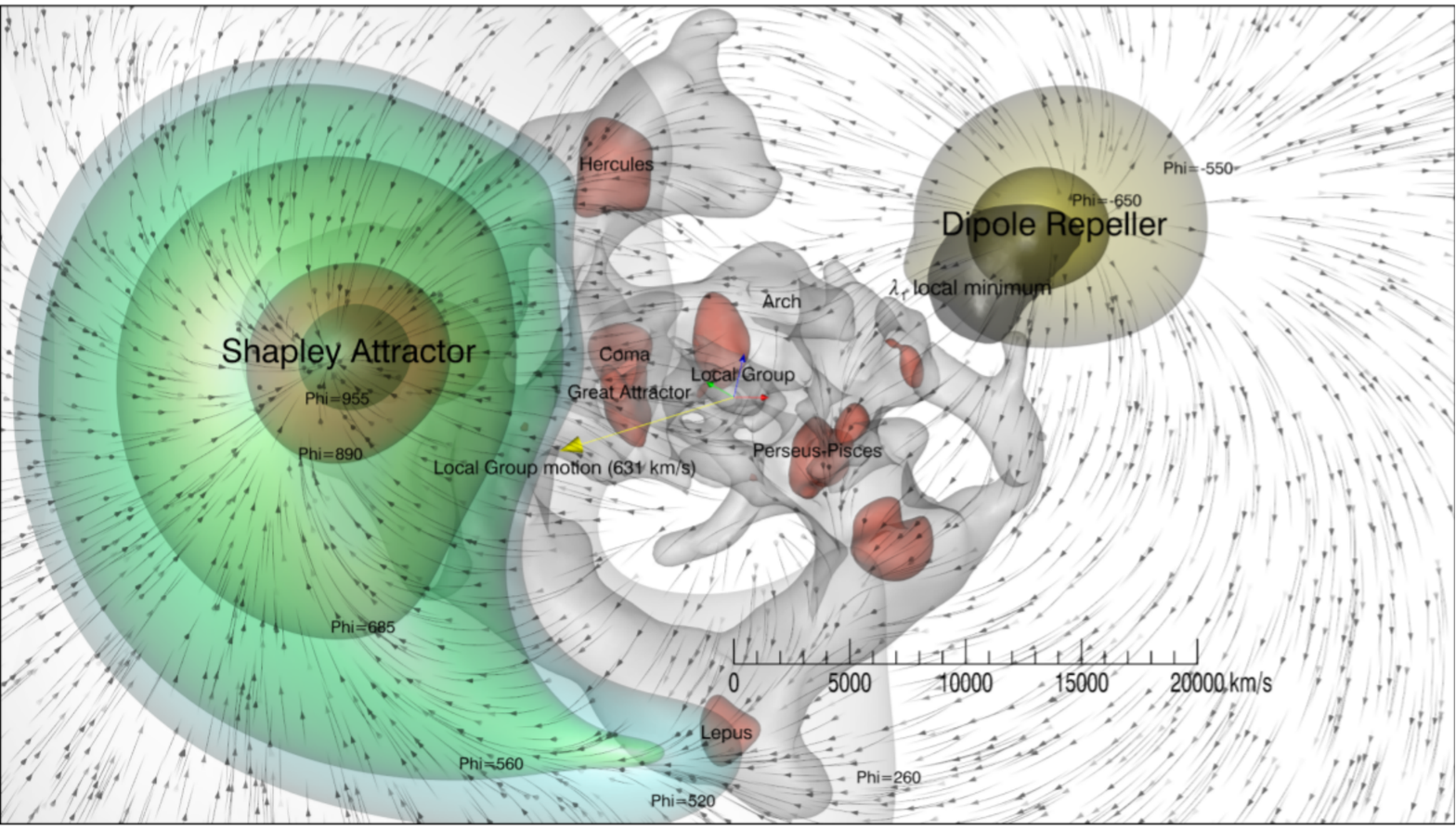}
\caption{A face-on view of a slice 6,000~\kms\  thick, normal to the directionn of the pointing vector $\hat{r}=(0.604, 0.720, -0.342)$. 
The scale can be inferred from the signpost made of three 2000 km/s-long arrows anchored at the origin of Supergalactic Coordinate System (SGX,SGY,SGZ),
with Red, Green, Blue arrows associated with the SGX, SGY, SGZ axes, respectively.
Three different elements of the flow are presented: mapping of the velocity field by means of streamlines (seeded randomly in the slice),
red and grey surfaces present the knots and filaments of the V-web, respectively and   equipotential surfaces are shown in green and yellow.
The potential surfaces are enclosing the  Dipole Repeller (in yellow) and the Shapley Attractor (in green) that dominate the flow. 
The yellow arrow indicates the direction of the CMB dipole ($gl=276^\circ$, $gb=30^\circ$).}
\end{figure}

Describing the gravitational dynamics in co-moving coordinates, by which the expansion of the universe is factored out, underdensities   apply a repulsive force and overdensities   an attractive one. 
We opt here to represent the vector  field by means of streamlines - the tangent of which is in the direction of the velocity vector and its colour represents the amplitude of the vector. 
The sources and  sinks of the streamlines are associated with   the   attractors and repellers of the large scale structure. 
These are closely associated with the voids and knots of the V-web. The voids (knots)  are regions of diverging (converging) flow, namely regions where the Hessian of the velocity potential is negative (positive) definite, yet these regions are in general moving with respect to the CMB frame of reference. The repellers and attractors are stationary voids and knots 
(respectively), hence they correspond to local extrema of the gravitational potential. (See figure 1 and the on-line video for a visualization of the velocity potential.)
Figure 2 shows a 3D visualization of the streamlines in a box of length 40,000~\kms\ centered on the Local Group. The  lines are seeded on a regular grid and  extend either to converge with a knot or exit the box.   All the flow lines of  the left plot of figure 2 either converge onto an attractor located roughly at  [$-$12,300, 7,400, $-$300]~\kms\ or cross out of the box. (Cartesian  Supergalactic   coordinates are assumed here.) The plot  uncovers the existence of a repeller at the upper right hand side of the box - a region from which  flow lines seem to diverge. Repellers  are best manifested by the  anti-flow, namely the negative of the velocity field.   The right plot of figure 2  depicts the convergence of the  streamlines of the anti-flow onto a repeller at  [11,000, $-$6,000, 10,000]~\kms. 
The WF reconstruction of the Cosmicflows-2 data detects a single attractor and a single repeller, the Shapley Attractor and the Dipole Repeller (hereafter the Attractor and the Repeller for brevity). (The accompanying video of the on-line version provides a further visualization.)

\begin{figure}
\includegraphics[width=1.0\textwidth,angle=00]{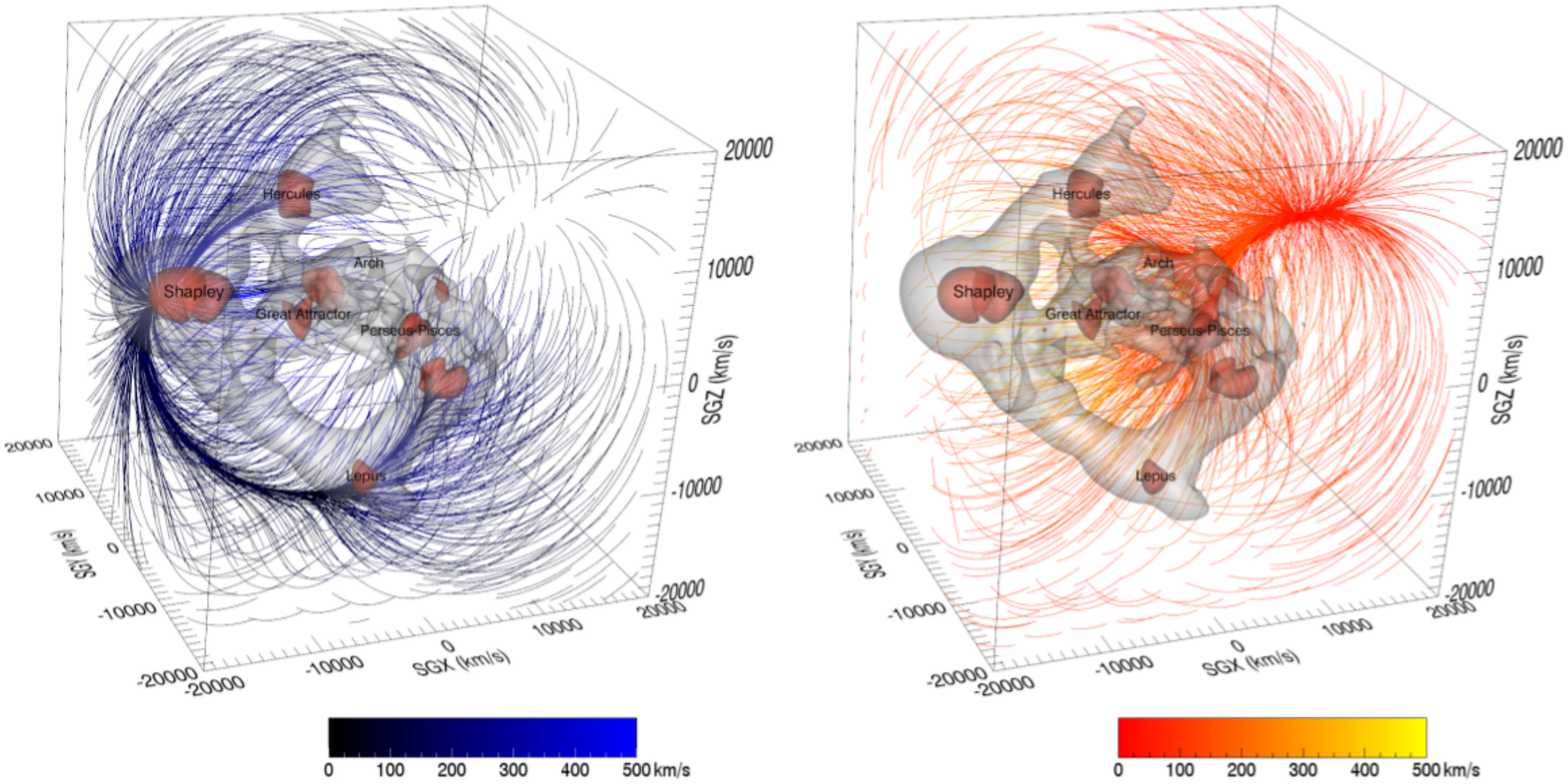}
\caption{A three dimensional (3D) view of the stream lines of the flow field (in black-blue, left panel) and of the anti-flow (in yellow-red, right panel). The stream lines are seeded on a regular grid and are coloured according to the magnitude of the velocity,  The flow stream lines clearly diverge from the Repeller and converge on the  Attractor. For the anti-flow the divergence and convergence are switching roles.  The knots and filaments of the V-web are shown  for reference. (For a 3D view look at the accompanying video, 00:56 - 01:28.)
}
\label{fig2}
\end{figure}

The WF recovers   the  Attractor and the Repeller  near  the edge of the  Cosmicflows-2 data. It is the long range correlation of the velocity field that renders the  imprint of the  Repeller and the  Attractor on the local flow. 
Multipole expansion provides a different insight into the nature of the local flow. Here the spherical top-hat weighted bulk flow  (i.e. dipole) and shear tensor (i.e. quadrupole moments) are evaluated at variable radius $R$\cite{1995ApJ...455...26J,1999ApJ...520..413Z,2001astro.ph..2190H,2010MNRAS.407.2328F}.
 A single attractor or repeller induces  a dipole and the expansion eigenvector of the shear tensor aligned with its direction and a degeneracy of the other two eigenvalues/eigenvectors. This is the telltale  signature a single dominant attractor or repeller. The local flow is not dominated by a single attractor or repeller. 
In the following we emphasize the directional aspects of the dipole and shear eigenvectors. The WF acts as an adaptive filter - where data is missing and/or noisy it suppresses more strongly the signal and the small scale structure.  Hence the contribution of the (better sampled) Attractor cannot be directly compared with the (extremely poorly sampled) Repeller. Yet, directions are robustly recovered by the WF. Figure 3 presents an aitoff projection of the following directions: 
1. the  Repeller; 
2. the  Attractor;
3. the CMB dipole and its anti-apex;
4. the bulk velocity of top-hat spheres of $R=(2,000, 3,000, ..., 15,000)$~\kms, \VbulkR, of the WF reconstructed flow field;
5. the three eigenvectors of the shear tensor ($\hat{\bf e}_i$,  i=1,2,3) of the WF field.
The figure shows the strong anti-alignment of the CMB dipole and the bulk velocity of spheres of radii smaller than 15,000~\kms\  with the  Repeller. Beyond that radius the bulk velocity looses its coherence in terms of direction, as the scatter in direction steadily increases.  The third eigenvector of the shear tensor ($\hat{\bf e}_3$), that reflects the direction of maximal expansion, is  aligned with the direction of the  Attractor out to $R=7,000$~\kms.  Figure  4   further presents the mean  and the scatter around the cosine of the angles formed between the bulk velocity and the  Repeller, $\mu_{\rm bulk}(R)=\cos({\bf V}_{\rm bulk},{\bf R}_{\rm GR}  )$,  and between $\hat{\bf e}_3$  and the  Attractor, $\mu_{\rm e3}(R)=\cos(\hat{\bf e}_3(R),{\bf R}_{\rm Shapley}  )$.

\begin{figure}
\label{fig3}
\includegraphics[width=1.\textwidth]{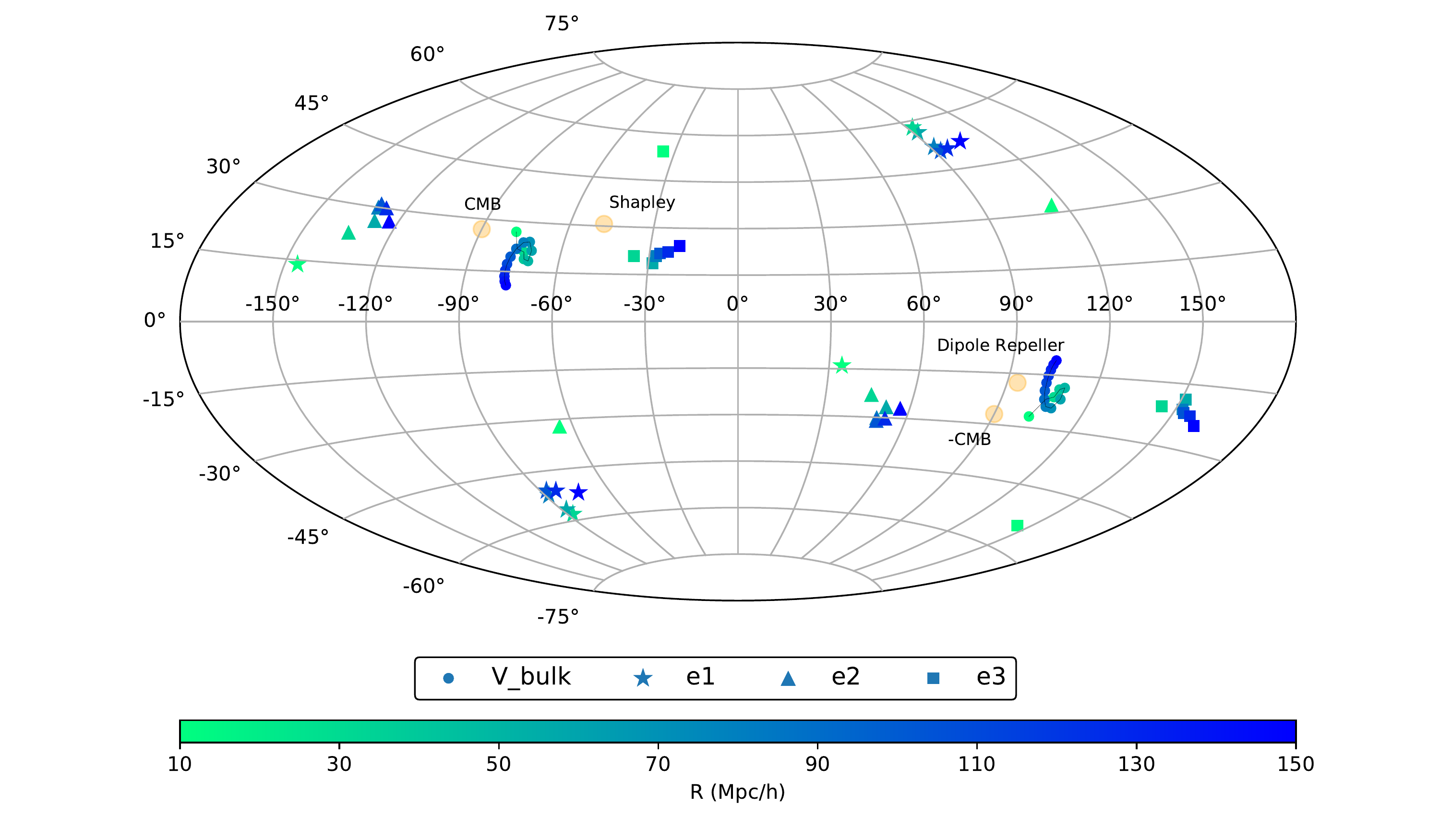}
\caption{Aitoff projection in  galactic coordinates of the principal structures and directions that characterize the flow: the  Dipole Repeller (GR), the Shapley Attractor, the  CMB dipole, the bulk velocity and the three eigenvectors of the velocity shear tensor. The latter two quantities are evaluated across spheres of radii ranging up to 30,000~\kms\ for the bulk velocity and 15,000 
\kms\ for the eigenvectors. The close (anti)alignment of the bulk velocity and the alignment of the $\hat{\bf e}_3$ eigenvector with the  Attractor are robustly manifested. 
}
\end{figure}

The close alignment of the bulk velocity with the Repeller out to  roughly $R=16,000$~\kms, where $\mu_{\rm bulk}(R)= -0.96 \pm    0.042$, provides the strongest support for the validity of the Repeller and for its dominant role in dictating the local flow. It is interesting to study the shear tensor. The expansion eigenvector is closely aligned with the direction to the Shapley Attractor out to $R\approx 7,000$~\kms\ - a direction which coincides with the Great Attractor, located at the bottom of the Laniakea basin of attraction\cite{2014Natur.513...71T} at (-4,700, 1,300, 500)~\kms. It is the combined mass distribution within the Laniakea and Shapley superclusters that dominates the tidal field - with the inverse cubic distance dependence of the tidal interaction tipping the balance towards the Laniakea/Great Attractor.

The main findings of the article are tested  against statistical and systematic uncertainties. 
There is no doubt about the existence  of the  Shapley Concentration  and therefore we focus our attention  mostly on the  Repeller.
The strong support for the existence of the Repeller comes not only from its close alignment of the bulk velocity but also the small scatter around the mean WF value, 
$\mu_{\rm bulk}(R)= -0.96 \pm    0.04$ for $R \approx 16,000$~\kms\ (figure 4). Assuming that the Repeller is the dominant structure that determines the direction of the bulk flow the scatter in $\mu_{\rm bulk}(R)$ can be translated to uncertainty in the position of the Repeller, $\Delta R_{\rm GR} \approx 4,500$~\kms\ (see Methods).
Next, a possible `edge of the data' effect is considered, driven by the concern that both the Attractor and the Repeller are located at the extremity of the Cosmicflows-2 data zone.  This issue has been addressed by restricting the full Cosmicflows data to spheres of radii 6,000, 8,000 and 10,000~\kms\ which contain 49\%, 67\% and 82\% of the full data, respectively.
The WF applied to these subsets of data recovers the  Repeller to within roughly  $\Delta R_{\rm GR}$.
Systematic errors, such as the ones introduce by the Malmquist bias, can introduce  systematic in- or outflows but the validity of the proposed morphology is supported by 
the reconstructed   back outflow from the Repeller.

The general picture that emerges here is of a complex flow that cannot be explained  by a simple toy model, yet the main structures that shape the local flow can be identified. 
The WF finds that the flow is dominated by a single attractor and a single repeller.
The dominance of the  Repeller is manifested by the fact that the CMB and the bulk velocity dipoles are all  strongly (anti)aligned with its direction. The  Repeller  pushes our local patch of the universe. The Repeller dominates the bulk flow out to   a distance of 16,000 $\pm$ 4,500~\kms and  the  Attractor dominates the shear  term out to roughly 7,000~\kms. In the language of multipole expansion the repeller dominates the dipole  and the attractor dominates the quadrupole moments. 
The role played  by the  Shapley Attractor is not surprising - the earlier findings on  influences beyond the Great Attractor\cite{1988ApJ...326...19L,1991Natur.350..391D,1989Natur.338..562S,1989Natur.342..251R,2006ApJ...645.1043K} suggested it. The existence of the Repeller was only very vaguely hinted before.
A study of the all-sky distribution of X-ray selected clusters uncovered a significant under-density of clusters in the northern hemisphere roughly 15,000~\kms\ away\cite{2006ApJ...645.1043K}. It suggested that this under-density  may be as significant as the overdensity of clusters in the southern hemisphere in inducing the local flow. 
Earlier examinations   of galaxy peculiar velocities found a north-south anisotropy in (galactic) y-component of the velocities\cite{2015MNRAS.447..132W} and that the  sources responsible for the bulk flow are at an effective distance  $>$30,000~\kms\  \cite{2010MNRAS.407.2328F}. Here, the source of the repulsion  is identified for the first time.
Arguably,  the  dominance of the Dipole Repeller over the Shapley Attractor   is the main novel and surprising finding of this study.  
The predicted position of the  Repeller is in a region that is yet poorly covered by existing redshift surveys. We predict the  Repeller to be associated with a void in the distribution of galaxies.

In the linear regime of gravitational instability repellers are as abundant and dominant as attractors. Yet observationally, repellers are  much harder to identify than attractors. The association of repellers  with underdensities  renders  them strongly deficient of galaxies, in general, and clusters of galaxies, in particular. The detection of voids by means of redshift surveys is challenging. Our use of peculiar velocities as tracers of the large scale structure overcomes that observational hindrance  and unveils the existence of the new structure we call the Dipole Repeller.

\section{methods}

{\bf Cosmicflows-2 dataset:} The present studies is based on the second release   catalogue of galaxy distances and peculiar velocities, Cosmicflows-2 \cite{2013AJ....146...86T},   that extends sparsely to recession velocities of 30,000~\kms\ (redshift $z\approx0.1$). It consists of  8,161 entries with high density of coverage inside 10,000~\kms. 
Here we used a grouped version of the Cosmicflows-2 data, in which all galaxies forming a group, of two or more, are merged to one data entry. The grouped Cosmicflows-2 data consists of 4885 entries. Six methodologies are used for distance estimation: 
Cepheid star pulsations, the luminosity terminus of stars at the tip of the red giant branch, surface brightness fluctuations of the ensemble of stars in elliptical galaxies, type Ia  supernovae, the  fundamental plane in luminosity, radius, and velocity dispersion of elliptical galaxies, and the Tully-Fisher correlation between the luminosities  and rotation rates of spiral galaxies.

{\bf Wiener filter and Constrained Realizations:} 
In the standard model of cosmology the linear  velocity field constitutes a Gaussian random vector field\cite{1980lssu.book.....P}. The Cosmicflows-2 dataset, as all other available velocity surveys, provides a sparse, incomplete, inhomogeneous and a very noisy sampling of the local flow. The Bayesian formalism of the WF and CRs provides the optimal methodology for the reconstruction (estimation) of the underlying velocity field and the associated uncertainties  in the linear regime\cite{1991ApJ...380L...5H,2009LNP...665..565H,1995ApJ...449..446Z,1999ApJ...520..413Z}. The WF/CRs reconstruction is based on an assumed prior cosmological model - the $\Lambda$CDM model  with the WMAP inferred cosmological parameters.  The current WF and CR fields are the ones reported  in our bulk velocity article\cite{2015MNRAS.449.4494H}. The results presented here are insensitive to the exact values of the  
$\Lambda$CDM  parameters, in particular to the differences between the WMAP and Planck parameters. 

{\bf Cosmic V-web:}
The cosmic web is defined here by the means of the V-web\cite{2012MNRAS.425.2049H}. 
The V-web is a mathematical model used to construct the cosmic web. The model starts with the continuous velocity field and the  velocity shear tensor evaluated for that field (equation \ref{eq-V-Sigma}). Consider a given point in space at which the  shear tensor is evaluated, and thereby its eigenvalues and eigenvectors. The so-called V-web is defined by a threshold values ($\lambda_{\rm th}$) - a free parameter which defines the web. The number of eigenvalues above  $\lambda_{\rm th}$ defines the web classification at that point - 0, 1, 2 or 3 corresponds to the point being a void, sheet, filament or knot.
The  normalized velocity shear tensor at a given grid cell, is defined by:
\begin{equation}
\Sigma_{\alpha\beta} = -  {1\over 2 H_0}   \big( \partial_\alpha v_\beta  +  \partial_\beta v_\alpha \big) 
\label{eq-V-Sigma}
\end{equation}
The standard definition of the velocity shear tensor is modified here by the Hubble constant ($H_0$) normalization, which makes it dimensionless. The minus sign is introduced so that a positive eigenvalue corresponds to a contraction. Eigenvalues are ordered by decreasing value, which implied that $\hat{e}_1$ points in the direction of maximum collapse and $\hat{e}_3$ points toward maximum expansion.
The V-web is defined by the effective resolution of the velocity field and by the value of the threshold. Here a Gaussian smoothing of $R_s=250$~\kms\ and $\lambda_{\rm th}=0.04$ are assumed. 

{\bf Multipole expansion of the flow:}
A first order expansion of a  potential (i.e. irrotational) velocity field, $\vec{v}(\vec{r})$, around a point labeled by ${\bf 0}$  yields,
\begin{equation}
v_\alpha(\vec{r}) \sim v_{0,\alpha} + (\partial_\beta\  v_\alpha)\  r_\beta =   v_{0,\alpha} - H_0\ \Sigma_{\alpha\beta}\ r_\beta,
\label{eq-1st}
\end{equation}
where $v_{0,\alpha}$ and  $\Sigma_{\alpha\beta} $ are evaluated at the point ${\bf 0}$.  
This expansion is equivalent to a  dipole and quadrupole expansion of the (velocity) potential. 
The flow in a sphere of radius $R$ is modelled here as the sum of a bulk flow, ${\bf V}{_{\rm bulk}}(R)$,  and a shear term, $ \Sigma_{\alpha\beta}(R)$, in the manner of equation \ref{eq-1st}. 
The parameters of the model, i.e. the bulk velocity vector and the symmetric tensor, are found by minimizing the quadratic residual between the model and the actual velocity field, with a spherical top-hat window function weighting.

{\bf Streamlines:} In the linear regime the flow is irrotational, namely it is a potential flow, and hence the velocity field can be written as a gradient of a scalar (velocity) potential, 
$\vec{v}=\nabla\phi_v(\vec{r})$. In this linear regime the peculiar velocity ($\vec{u}$) and gravitational field  ($\vec{g}$)  are simply related by
$\vec{u}={2\over 3} {f(\Omega_m, \Omega_\Lambda) \over  H \Omega_m} \vec{g} $  where $\Omega_m, \Omega_\Lambda$ and  $H$ are the time dependent cosmological matter and dark energy density parameters and Hubble's constant, respectively. The velocity and gravitational potential are similarly related. Inspired by the similarity between the gravitational potential in linear theory, hence also the velocity potential, to the electrical potential in electro-statistics, we present the flow field by field lines which we call here stream or flow lines. A trajectory of a streamline, $\vec{l}(s)$ where $s$ is the line parameter, is calculated by integrating the line equation, $d \vec{r}(s) = \vec{v}(\vec{r}(s)) d s$. The numerical calculation of a streamline  involves the determination of the seeds of the stream lines and the number of integration   steps. For small number of integrations steps and a regular grid of seeds the streamlines resemble velocity arrows. For a large number of steps the flow and anti-flow lines are either trapped by attractors or repellers or leave the box. It should be emphasized here that the streamlines are a graphical mean for the presentation of a vector field and do not represent trajectories of objects.

{\bf Video:} 
The video  (http://vimeo.com/pomarede/dipolerepeller or http://irfu.cea.fr/dipolerepeller) commences with the presentation of the large scale structure by means of the surfaces that define the filaments and knots of the V-web. The motion of the Local Group with respect to the CMB is displayed by a yellow arrow. The growth of  streamlines through integration steps from a regular array of seeds are illustrated. The fully developed streamlines clearly shows a pattern dominated by a single repeller and a single attractor. The growth of the streamlines of the anti-flow is shown in a similar manner - the anti-flow is being repelled by the  Attractor and attracted by the  Repeller. A different presentation of the flow and anti-flow streamlines is obtained by confining the seeds of the flow lines to the neighbourhood of the   Repeller  (in blue-black) and those of the anti-flow to the vicinity of the  Attractor (in orange-red). The attractive flow develops mainly in a plane containing the major over-densities: Perseus-Pisces, Lepus and Hercules.
The repulsive anti-flow develops in the orthogonal plane that corresponds roughly to the supergalactic equator. The video reveals the almost exact alignment of the Local Group velocity vector with the Dipole Repeller and the much poorer alignment with the Shapley Attractor. The  Repeller and  Attractor constitute local maxima and minima of the gravitational potential.

{\bf Uncertainties assessment:}
The probability distribution  of the alignment of the bulk velocity and the eigenvectors of the shear tensor is sampled  by means of an ensemble of 20 CRs, constrained by the Cosmicflows-2 data and evaluated within the WMAP parameters of the  \LCDM\ model. The CRs are evaluated on a grid of size $256^3$ spanning a box of 256,000~\kms\ on its side.  The bulk velocity and the velocity shear tensor are obtained by a convolution of the velocity field with a spherical top-hat window of radius $R$ and are evaluated at the center of the box, i.e. the location of the Local Group. Figure 4 presents the alignment of the bulk velocity and the 3rd eigenvector of the shear tensor with the  Repeller and the  Attractor respectively over a range of radii of $R=(2,000, 3,000, ... 30,000)$~\kms. The uncertainty in $\mu_{\rm bulk}(R)$  is used  to assess the uncertainty in the position of the Repeller. At $R \approx 16,000$~\kms\ we find   $\mu_{\rm bulk}(R)= -0.96 \pm    0.04$. Assuming the Repeller to be responsible for the direction of the bulk velocity the uncertainty in    $\mu_{\rm bulk}$ is translated   to a projected distance (at a distance of 16,000~\kms) we find $\Delta R_{\rm GR} \approx 4,500$~\kms.
The  uncertainty in $\mu_{\rm bulk}$   changes significantly over the range $R=$ 11,000 to 20,000~\kms. This again translates to an uncertainty in the radial position of roughly 4,500~\kms.

Next  a possible `edge of the data' systematic effect is considered.  To meet this end  the full Cosmicflows data has been trimmed in  spheres of radii 6,000, 8,000 and 10,000~\kms\ which contain 49\%, 67\% and 82\% of the full data, respectively. The WF has been applied to these subsets of data and the resulting (anti)flow field has been compared with that of the full data (figure 5). The overall structure of the flow fields of the subsamples  follow that of the full Cosmicflows-2 data. 
The anti-flow converges into two repellers in the 6,000~\kms\ case and into single repellers for the 8,000 and 10,000~\kms\ cases. Table 1 provides the location and distances from the Local Group and the Repeller for each case. It also provides the fraction of data contained in the each subsample. It is remarkable that with less than half the data (6,000~\kms\ case)  the WF recovers the general (anti)flow towards the general position of the Repeller. Taking the mean position of the two repellers it is found to be a mere 2,500~\kms\ from the Repeller. Single repellers are found  for the larger subsamples that consist of 67\% and 82\% of the data  at distances of 4,300 and 500~\kms\ from the Repeller, respectively. We conclude that the Repeller is not  a fictitious structure  induced by an 'edge of the data' effect.

\bigskip

\begin{figure}
\label{fig4}
\includegraphics[width=1.0\textwidth]{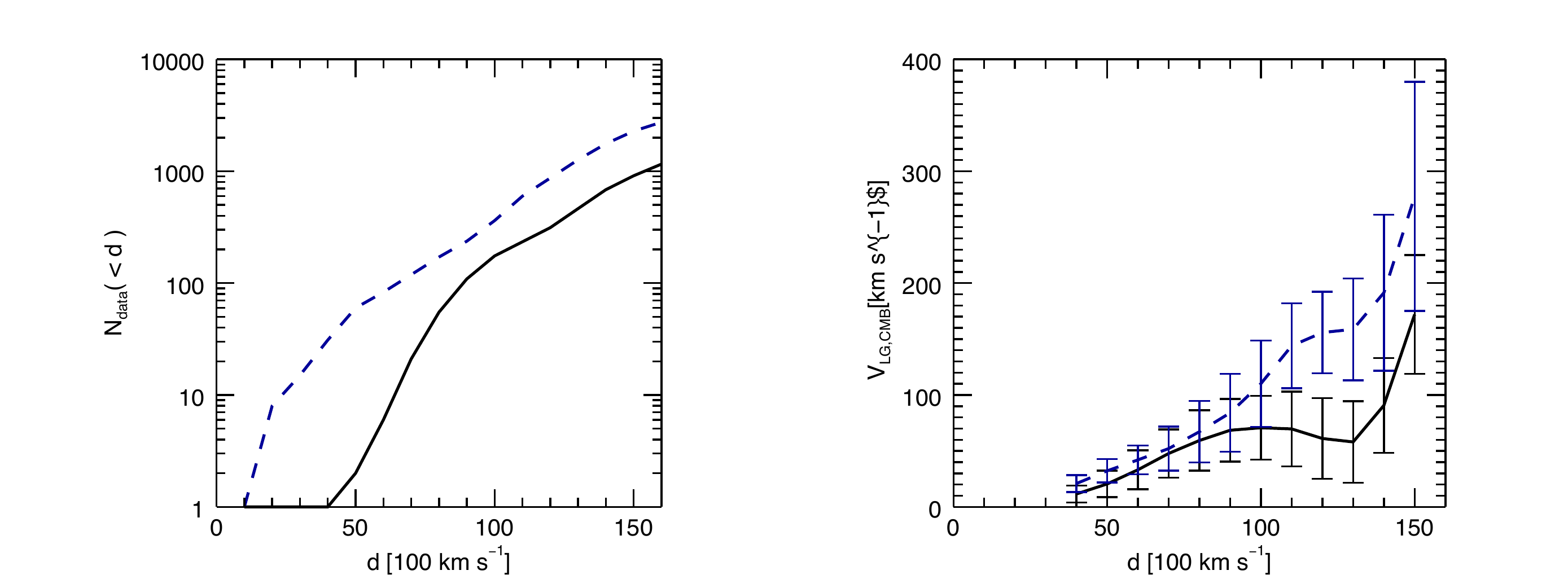}
\caption{Alignment of the bulk velocity with the direction of the  Repeller in the lower set of curves and of the $\hat{e}_3$ eigenvector with the  Attractor in the upper set of curves. The alignment is presented by the mean (solid lines) and the mean $\pm$ standard deviation (dashed lines), where the statistics are calculated over an ensemble of 20 constrained realizations. The alignment with the Attractor is expressed by $\mu_{\rm e1}(R)$ (upper curve) and with the Repeller by $\mu_{\rm bulk}(R)$ (lower curve).  The downward black arrow and the upward blue arrow indicate the distances of the  Repeller and the  Attractor, respectively, from the Local Group.
}
\end{figure}

\begin{figure}
\label{fig5}
\includegraphics[width=1.0\textwidth]{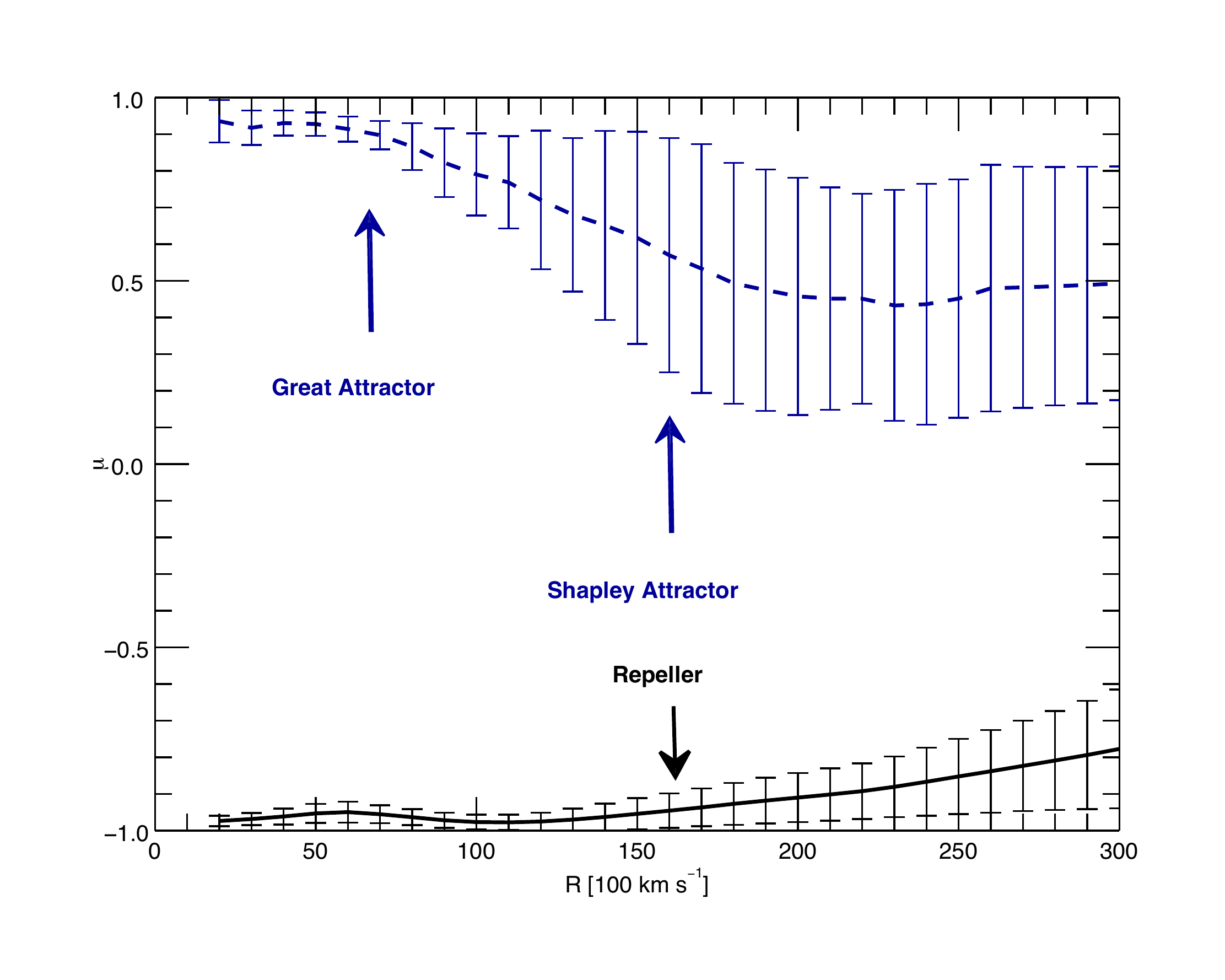}
\caption{The anti-flow streamlines of the WF reconstruction of the three trimmed Cosmicflows-2 data: The $R=6,000$ (left panel), $8,000$ (middle panel) and $10,000$\~\kms (right panel). 
These plots need to be compared with the equivalent plot made out  of the full dataset (right panel of figure \ref{fig2}). In the case of the shallowest  data ($R=6,000$~\kms) the repeller breaks into two separate repellers. For the other datasets the flow convergence to a single repeller. Table \ref{table1} presents the location of the repellers of the sub-sample, their distances from the Local Group and from the Repeller. 
}
\end{figure}

\begin{table}
\begin{tabular}{ccccc}
\hline
distance cut [\kms]  &  \% of data  &  [SGX, SGY, SGZ]  [\kms] & R [\kms]  & $d_{GR}$ [\kms]  \\
\hline
6,000      &        49      &       [12,500,    2,600,         9,800]     &       16,100   &  8,700    \\
              &                   &      [10,000,    -11,600,    10,500]      &      18,500   &  5,700    \\
8,000      &        67       &      [10,600,     -5,800,     10,000]      &       13,700    &   4,300    \\
10,000    &        82       &      [10,900,     -2,100,       8,100]     &       15,600   &   500    \\
full          &      100       &      [11,000,    -6,000,      10,000]    &       16,000   &   0  \\
\hline
\end{tabular}
\caption{Cosmicflows-2 subsamples: The location, distance ($R$) and the distance from the  Repeller ($d_{GR}$) of the repeller found for the different distance cuts of the data (first column). 
Two repellers are identified for the 6,000~\kms\ subsample and the first two raws present their locations and distances. The mean position of these two repellers is located at a distance of $d_{GR}$=2,500~\kms. 
}
\label{table1}
\end{table}


\newpage

\bibliography{biblio_GR}

 Acknowledgments : Jenny Sorce and Stefan Gottloeber are gratefully acknowledged for valuable discussions
 and Alexandra Dupuy for her help in preparing Figure 3.
 We express our gratitude to K. Bowles and S. Thompson for the narration in the supplementary video. Support has been provided by the Israel Science Foundation (1013/12), the Institut Universitaire de France, the US National Science Foundation, Space Telescope Science Institute for observations with Hubble Space Telescope, the Jet Propulsion Lab for observations with Spitzer Space Telescope and NASA for analysis of data from the Wide-field Infrared Survey Explorer.\\
 
Support has been provided by the Israel Science Foundation (1013/12), the Institut Universitaire de France, the US National Science Foundation, Space Telescope Science Institute for observations with Hubble Space Telescope, the Jet Propulsion Lab for observations with Spitzer Space Telescope and NASA for analysis of data from the Wide-field Infrared Survey Explorer.
The authors declare that they have no
competing financial interests.
Correspondence and requests for materials
should be addressed to Y.H.~(email: hoffman@huji.ac.il).

{\parindent 0pt
{\bf Data availability statement.} The data that support the plots within this paper and other findings of this study are available from the corresponding author upon reasonable request. It is also available freely at the Extragalactic Distance Database: edd.ifa.hawaii.edu
and through the NED interface. Use of the data and flow model must cite this article.\\
}

\end{document}